\documentstyle[sprocl]{article}
\def\be{\begin{equation}}
\def\ee{\end{equation}}
\def\bea{\begin{eqnarray}}
\def\eea{\end{eqnarray}}
\def\R{{\rm I\!R}}

\def\Zop{{\rm Z\!\!Z}}
\def\H{{\cal H}}
\def\F{{\cal F}}
\def\K{{\cal K}}
\def\S{{\cal S}}
\def\M{{\cal M}}
\def\SU{ {\rm SU}(1,1) }
\def\cspace{\hskip 1.5pt}
\begin{document}
\phantom{a}
\vskip -5em
\rightline{DAMTP-1998-172}
\rightline{hep-th/9812252}
\vskip 2em
\title{THE NO-GHOST THEOREM AND STRINGS ON AdS${}_{\bf 3}$\cspace\footnote{
Talk presented by JME} }
\author{ J.M.~EVANS, M.R.~GABERDIEL, M.J.~PERRY}
\address{DAMTP, University of Cambridge\\
Silver Street, Cambridge CB3 9EW, UK}
%
%

\maketitle
\abstracts{
A brief review of string theory on group manifolds is given,
and comparisons are then drawn between Minkowski space, SU(2), and 
${\rm SU}(1,1) = {\rm AdS}_{3}$.
The proof of the no-ghost theorem is 
outlined, assuming a certain restriction on the representation
content for bosonic and fermionic strings on $\SU$.
Some possible connections with the AdS/CFT correspondence are mentioned.}

\section{Introduction}

The no-ghost theorem for strings in flat space\cspace\cite{GT,B} 
is such a long-standing and fundamental result in the subject 
that there is some danger of taking it for granted.
In the simplest version one considers traditional covariant 
quantization of the bosonic string in Minkowski space. 
The theorem states that the Fock space inner-product is
positive-semi-definite on the subspace of {\it physical states}, defined
as those that satisfy the {\it mass-shell\/} and {\it Virasoro
primary\/}  conditions: 
\be
\label{ms}
(L_0 - 1 ) | \psi \rangle = 0 \, , \qquad 
L_n | \psi \rangle = 0 \qquad n>0 
\ee
respectively, where $L_n$ obey the Virasoro algebra with $c=26$.
Since string theory is a theory of gravity, however, we should regard
Minkowski space as merely a particular choice of background.
But it is far from clear how the no-ghost theorem and its proof 
can be generalized when one moves from flat space to 
more general background spacetimes. 

The standard criteria for consistency of a target spacetime in
perturbative string theory is that the
world-sheet sigma-model which it defines should be quantum conformally
invariant, as
given by the vanishing of appropriate $\beta$-functions for the
metric, dilaton, and other background fields.  But it is not
hard to see that this approach is sometimes insensitive to 
important aspects of the theory.
By these criteria alone, a flat spacetime 
with 13 time-like and 13 space-like directions would seem to be a
perfectly consistent
background for the bosonic string, with $c=26$, and yet this model
certainly contains ghosts---i.e.~physical states (in the sense of
(\ref{ms})) which have negative norm.

It seems reasonable to restrict attention to backgrounds with a {\it
single\/} time-like direction (though there have recently been more
radical suggestions\cspace\cite{Hull}).  The original proofs of the
no-ghost theorem\cspace\cite{GT,B} are easily applied to critical
strings moving in target spaces of the form $\R^{d-1,1} \! \times
{\cal M}$ where $2 \leq d \leq 26$ and ${\cal M}$ corresponds to a
unitary CFT of appropriate central charge.  If on the other hand we
consider a background whose geometry involves a time-like direction in
a non-trivial way, then issues such as unitarity and the absence of
ghosts must be scrutinized very carefully.

To examine such questions it is natural to turn to the simplest string
models which one can hope to solve exactly, namely those for which the
target spaces are group manifolds.\cite{GW} Requiring a single
time-like direction leads unavoidably to the non-compact group ${\rm
SU}(1,1) = {\rm SL} (2,\R)$ or its covering group.  This was exactly
the route followed by Balog et al.\cite{BOFW} and it led them to some
unexpected and superficially discouraging conclusions.  They found
that although classical string propagation on ${\rm SU} (1,1)$ is
consistent and causal, there are nevertheless negative-norm physical
states in the quantum theory.

There have been a number of suggestions in the intervening decade as
to how one might sensibly interpret $\SU$ string theory in the face of
these facts.  One proposal\cspace\cite{Petr,Moh} is to restrict the
allowed Kac-Moody representations in terms of their spin and the
level, in imitation of the unitarity condition for compact Kac-Moody
algebras.  This was pursued by Hwang et al.\cite{Hwang1,Hwang2} and it
is also the point of view adopted by us.\cite{EGP} An alternative
approach\cspace\cite{Bars,Satoh} involves an apparently different
definition of the quantum theory in terms of free-field-like Fock
spaces, with the introduction of additional singularities in the
Kac-Moody currents.  It is not clear how these approaches are related,
if at all, and although this is an interesting question it is not one
we are able to comment on here in any more detail.

The issues we have highlighted become all the more relevant in view of
some celebrated developments that have taken place in the past year.
It is by now a famous conjecture that there is a duality relating
supergravity (as the low energy limit of string theory) on
$n$-dimensional anti-de Sitter space, AdS${}_n$, and a conformal field
theory that lives on its $(n{-}1)$-dimensional boundary.\cite{M} A
specific example\cspace\cite{M,MS} involves type IIB string theory on
${\rm AdS}_3 \times S^3 \times \M^4$ (where $\M^4$ is either K3 or
$T^4$) which is conjectured to be dual to a two-dimensional conformal
field theory with target manifold a symmetric product of a number of
copies of $\M^4$. This example is of special interest as both partners
of the dual pair are simple enough to be analyzed explicitly and it
should therefore be possible to subject the proposal to non-trivial
tests.  On the string theory side, for instance, we have ${\rm AdS}_3
= {\rm SL}( 2,\R) = {\rm SU}(1,1)$ and $S^3 = {\rm SU}(2)$, and since
these are both group manifolds we should be able to determine the
spectrum of string states exactly.  The AdS/CFT correspondence
therefore provides an additional specific and powerful motivation for
clarifying the status of string theory on ${\rm SU}(1,1)$.

Here we give a brief summary of strings on group
manifolds\cspace\cite{GW} with a view to comparing and contrasting the
target space $\SU$ with Minkowski space and with SU(2).  We then
outline the arguments used to prove the no-ghost theorem, following
our recent work\cspace\cite{EGP} for bosonic and fermionic $\SU$
strings. This builds on the earlier references cited above, but we
hope that it also resolves some of the confusion that seems to have
persisted in the literature for a number of years. Finally we comment
on an intriguing prediction of the AdS/CFT conjecture and how this
might be reflected in the traditional string picture.  Since this work
was presented, there have been a number of very interesting
developments which attempt to understand string theory on AdS$_3$ in
more detail.\cite{dB,GKS,KLL,dBORT}

\section{ WZW models and strings}

A string moving on a group manifold $G$ is described by a world-sheet
WZW conformal field theory with the level $k$ proportional to the
string tension. More precisely, the background is the bi-invariant
metric on the group and an anti-symmetric tensor field strength
proportional to the parallelizing torsion.  The space of string states
is constructed in terms of the Kac-Moody (KM) algebra for $G$ with
generators $J^a_n$ obeying
\be
[J^a_m,J^b_n] = i f^{ab}{}_{\!c} J^c_{m+n} + k m \eta^{ab}
\delta_{m+n} 
\ee
where $\eta^{ab}$ is the Killing form 
(used to raise and lower indices) and $f^{ab}{}_{\! c}$ are 
the structure constants. 
This should be regarded as a non-abelian generalization of 
a set of harmonic oscillators.

The Sugawara expression for the Virasoro generators is
\be
\label{sug}
L_n = {1 \over 2k + Q_{\rm ad} } 
\sum_\ell \eta_{ab} : J^a_\ell J^b_{n-\ell} :
\ee
where the normal ordering procedure sends 
$J^a_n$ to the right and left for $n>0$ and $n<0$ respectively. 
It follows that 
\be
\label{vir}
[L_m,L_n]=(m-n) L_{m+n} + {c \over 12} m (m^2 -1) \delta_{m+n} 
\, , \quad \
[L_n,J^a_m] = -m J^a_{n+m} 
\ee
where the central charge is 
\be
\label{bosc}
c= {k \over k+ {1\over 2} Q_{\rm ad}} \, {\rm dim} G \, .
\ee
At grade zero, 
the KM algebra contains a copy of the finite-dimensional Lie algebra 
$G$ with generators $J^a_0$.
We write $Q = \eta_{ab} J^a_0 J^b_0$ for the quadratic Casimir
of this subalgebra, and $Q_{\rm ad}$ appearing above is its
value in the adjoint representation. 

To construct string states we choose a unitary {\it base
representation\/} $\tau$ of the zero-grade subalgebra 
acting on states $| i \rangle$. We declare 
\be 
\label{base}
J^a_0 | i \rangle = \tau^a_{ij} | j \rangle \, , \qquad
J_{n}^{a} |i \rangle \quad n > 0 
\ee
and the string Fock space is spanned at each grade $N$ by states 
\be 
\label{fock}
| \psi \rangle = J_{-n_1}^{a_1} \ldots J_{-n_r}^{a_r} 
| i \rangle \quad \mbox{with   }
N = \sum_k n_k \ .
\ee
Note that on such a state of grade $N$ we have 
\be 
\label{Lo}
L_0 | \psi \rangle = \left ( N + {Q_\tau \over 2k + Q_{\rm ad}} \right ) |
\psi \rangle \ . 
\ee

The superstring in the RNS formalism 
is constructed in a similar fashion from a super WZW model with target
space $G$.
In addition to the KM currents there are fermionic superpartners 
in the adjoint representation.
These can be decoupled by a re-definition of the currents, however,
resulting in a bosonic Kac-Moody algebra with shifted level
$\hat k = k - {1 \over 2} Q_{\rm ad}$ which commutes with the
fermions. It follows that the super-Virasoro algebra 
has (supersymmetric) central charge
\be
\label{superc}
\hat c = \left ( 1 - {Q_{\rm ad} \over 3k} \right ) {\rm dim} G \ .
\ee
(As usual $c = 3\hat c /2$.)

So far we have said nothing about the nature of the group $G$.  If $G$
is compact and semi-simple then $k$ must be quantized to ensure
quantum consistency.  With our normalization this means $k \in {1
\over 2} \Zop$.  It is also well-known that in this case the KM
representation constructed via (\ref{base}) and (\ref{fock}) is
unitary if and only if the highest weight of the base representation
$\tau$ and the highest root of $G$ have an inner-product that is
bounded by the level $k$. (In the supersymmetric case we replace $k
\rightarrow \hat k$.)

\subsection{ $G = \R^{d-1,1}$ }
This is the string in $d$-dimensional Minkowski space, and we may set
$k=1$ without loss of generality. The KM currents become the usual
string oscillators $J^a_n \rightarrow \alpha^a_n$, and the zero-grade
subalgebra is generated by the commuting momentum operators $J^a_0
\rightarrow p^a$.  Base representations are simply states of definite
momentum $| p^a \rangle$.  The Virasoro primary conditions determine
the physical polarizations of states, while the mass-shell condition
in conjunction with (\ref{Lo}) implies that for physical states
$(mass)^2 = - p^2 \sim N - 1$.  There are finitely many physical
states at each possible grade $N$.

\subsection{ $G = {\rm SU(2)} = S^3$ }
Since the group is compact, $k$ must be a half-integer, and other
relevant data are 
$\eta_{ab} = {\rm diag}(+1,+1,+1)$, $f_{abc} = \varepsilon_{abc}$ and 
$Q_{{\rm ad}} = 2$.
The resulting central charges for the bosonic and supersymmetric
theories are then
\be
\label{cc}
c = {3k \over k+1} \, , \qquad \hat c = 3 - {2 \over k} \ .
\ee
The base representations are the standard unitary representations of 
SU(2), with states $| j , m \rangle$ labelled by their eigenvalues: 
$$ 
Q | j , m \rangle = j(j+1) | j , m\rangle \ ,  \qquad 
J^3_0 | j , m \rangle = m | j , m\rangle \ .
$$
The condition for a unitary representation of the KM algebra 
is 
\be
\label{csl}
j \leq k 
\ee
for the bosonic theory; in the supersymmetric theory we simply replace
$k \rightarrow \hat k = k-1$.
The physical state conditions are not 
relevant to this model if it is taken in isolation
because there is no time-like direction in the target space.
It can of course appear as a factor in some larger target space,
as we shall see below.

\subsection{ $G = {\rm SU}(1,1) = {\rm AdS}_3$ }
Since the target manifold is non-compact, $k$ is not quantized and can 
apparently be chosen at will. The basic data are
$\eta_{ab} = {\rm diag}(+1,+1,-1)$, $f_{abc} = \varepsilon_{abc}$ and
$Q_{{\rm ad}} = -2$. 
The values of the bosonic and fermionic central charges are then
\be
\label{ncc}
\qquad c = {3k \over k-1} \, , \qquad \hat c = 3 + {2 \over k} \ .
\ee

As in the compact case, base representations of the zero-grade 
$\SU$ algebra are written $| j , m \rangle$ corresponding to 
eigenvalues 
\be
Q |j,m\rangle = - j (j+1) |j,m\rangle \ , \qquad
J^3_0 |j,m\rangle = m |j,m\rangle \,.
\ee
Since $\SU$ is non-compact, however,
its unitary representations are necessarily infinite-dimensional,
excepting the trivial representation consisting of the
single state with $j = m = 0$.

There are several types of non-trivial 
unitary representations of $\SU$. 
The {\it discrete\/} representations are: 
\be
D_j^{\pm} = \{ \, |j, \mp m\rangle \, : \, m = j, j{-}1, j{-}2, \ldots \,
\} \quad \mbox{ with } \quad J_0^\mp | j , \mp j \rangle = 0 
\ee
and they exist only for the values  
$j = -1/2,-1,-3/2,\ldots ,$ which explains their name. 
In contrast there are also {\it continuous} representations,
called {\it principal\/}, for which $j = -1/2 + i \kappa$ ($\kappa$
real), and {\it exceptional\/}, for which $-1/2 \leq j < 0$,
with the values of $m$ quantized in either case.
These continuous representations 
have no highest- or lowest-weight states.
Notice also that for both kinds of continuous 
representation $j(j+1) < 0$.

These are the only unitary representations of the group 
$\SU$ itself. If we pass to the covering group, however, then
there are more general representations of type $D_j^{\pm}$ 
in which $j$ and $m$ need not be half-integral
(although the allowed values of $m$ within any irreducible
representation always differ by integers) and similarly there are
additional continuous representations with the same ranges of values 
for $j$ but with more general values allowed for $m$.
The group manifold of $\SU$ is topologically
$\R^2 \times S^1$ (with the compact direction being time-like) and
this is responsible for the quantisation of $m$ in units of half
integers. By contrast, the simply-connected covering group is
topologically $\R^3$ and so there is no quantization of $m$ in this
case.   Our treatment of the no-ghost theorem\cspace\cite{EGP} 
applies equally well to either $\SU$ or its covering space.

\subsection{ ${\rm AdS}_3$ and the spin-level restriction }

The minus sign in the metric indicates that the generator 
$J^3_n$ plays the role of a time-like oscillator, creating 
negative norm states. The presence of such states in a covariantly
quantized string theory is hardly surprising.
The problem for the $\SU$ theory 
is that some of these negative norm states also satisfy the 
physical state conditions (\ref{ms}) and so would seem to be part of
the physical spectrum.
These difficulties only arise for states built on base
representations belonging to the discrete series $D_j^\pm$, however.
For all other base representations the mass-shell
condition implies that physical states occur only at grade zero,
and hence their norms are positive.

To exclude the unwanted ghost states, it was
suggested\cspace\cite{Petr,Moh} that a restriction should be imposed 
on the spin $j$ of the discrete series base representation in terms of
the level $k$ by analogy with the compact case (\ref{csl}). 
For the bosonic $\SU$ theory the condition is 
\be 
\label{ncsl}
| j | \leq k \  
\ee
while for the fermionic theory we again replace 
$k \rightarrow \hat k = k + 1$.
It is easy to check by some explicit calculations that this has 
the desired effect on states at low-lying grades.
Moreover, it can be shown to imply a completely ghost-free spectrum, 
as we shall sketch below.

The adoption of the above condition may look superficially natural,
but after more careful consideration it becomes clear that something
rather subtle is occurring and that the condition is working quite
differently in the compact and non-compact cases.  In the SU(2) model
the restriction (\ref{csl}) guarantees unitary representations of the
entire Kac-Moody algebra.  In the non-compact case it is impossible to
construct unitary representations of the Kac-Moody algebra by
(\ref{base}) and (\ref{fock}), whether one imposes (\ref{ncsl}) or
not.  Instead (\ref{ncsl}) forces physical states within these
non-unitary representations to have positive norm.

The condition (\ref{ncsl}) also has unusual 
implications for the spectrum of the theory.
It can easily be combined with the mass-shell condition (\ref{ms}) to
give  
\be
\label{restriction}
N < 1 + {k \over 2} 
\ee
(which holds for both the bosonic and fermionic cases).
Thus for a fixed level $k$ the physical states in the theory
can only arise at a finite number of grades. There are
infinitely many physical states at every allowed grade, however, since
the unitary base representations of ${\rm SU}(1,1)$ are
infinite-dimensional. 
This should be contrasted with the string in flat space,
which has finitely many physical
states at each of infinitely many grades.

\section{ The no-ghost theorem }

\subsection{ Skeleton of the proof} 

The approach of Goddard and Thorn\cspace\cite{GT} can be conveniently
divided into three steps, each of which must be established in order 
to show that a given model is ghost free.
(We consider only bosonic strings for simplicity.)

Let $\H$ be the string Fock space and $\F$ the subspace
of {\it transverse\/} states defined by
\be
L_n |f \rangle = K_n | f \rangle = 0 \quad n > 0
\ee
where $K_n$ are components of some chosen spin-1 current. 
We first require:
\hfil \break
{\bf (a)} The set of states 
\be 
\label{basis}
|\, \{\lambda,\mu\} \, f\, \rangle= L_{-1}^{\lambda_1} \cdots
L_{-m}^{\lambda_m} K_{-1}^{\mu_1} \cdots K_{-m}^{\mu_m}
|f\rangle \, , \quad | f \rangle \in \F 
\ee
is a basis for $\H$.
\hfil \break
We then define a subspace $\K$ to be the span of those states with 
$\lambda_i = 0$ and a 
subspace $\S$ of {\it spurious} states with at least one $\lambda_i \neq 0$.
Clearly these subspaces are complementary: $\H = \K + \S$.
Notice that a spurious state is orthogonal to any physical state. 
It is also easy to show that if $c=26$ 
the operators $L_n$ $(n > 0)$ map each of these subspaces into themselves:  
$\S \rightarrow \S$ and $\K \rightarrow \K$.
This has the important consequence that if 
$| \psi \rangle = | k \rangle + | s \rangle$ is Virasoro
primary, then $| k \rangle $ and $| s \rangle$ are separately 
Virasoro primary.
We must then establish:
\hfil \break
{\bf (b)} If $ | k \rangle \in \K$ is physical, then it is
transverse, $| k \rangle \in \F$.
\hfil \break
{\bf (c)} The inner-product on the transverse space $\F$ is positive
definite.  
\hfil \break
It is now easy to see how to prove the theorem.
If $| \psi \rangle = | k \rangle + | s \rangle$ is
physical, then $|k \rangle$ is transverse and 
$ \| \, | \psi \rangle \, \|^2 = \| \, | k \rangle \, \|^2 \geq 0$, 
which is the desired result.

In the original proof for the string in flat space, 
important simplifications were achieved by choosing $K_n$ to 
correspond to a light-like direction in Minkowski space.
The proof of {\bf (a)} could then be carried out by ordering the
states at each grade in an intelligent way,
while steps {\bf (b)} and {\bf (c)} become essentially trivial.
For ${\rm SU}(1,1)$ it is tempting to choose $K_n$ along a light-like
direction in a similar fashion\cspace\cite{Hwang1}
but unfortunately the same simplifications do not occur.
It is possible that the proof could be completed along these lines,
but there are serious technical obstacles.

An alternative approach is to choose $K_n$ corresponding to a time-like
direction.\cite{Hwang2,EGP} 
This means that steps {\bf (a)} and {\bf (b)} above
become considerably more difficult to establish, even for the string 
in flat space.
Fortunately, we can make use of the powerful general approach to CFT that 
has been developed in the meantime, and in particular the Kac
determinant formula. This allows us to establish steps
{\bf (a)} and {\bf (b)}
either in flat space, or for the target space $\SU$ with the
restriction (\ref{ncsl}).

Even for $K_n$ time-like, the final step {\bf (c)} is 
immediate in flat space, but it is highly non-trivial for the case of 
$\SU$ and it is here that the restriction (\ref{ncsl})
again enters crucially.
In fact the necessary result had already been established 
in a rather different context by Dixon et al.\cite{DPL}.
They showed that precisely for the restricted set of KM
representations given by (\ref{ncsl}) the coset 
$\SU$/U(1) has a positive-definite inner-product, where 
the U(1) corresponds to a time-like direction.
This completes the proof.

\subsection{ Ghost-free models }

There are now a number of models of the form
${\rm SU}(1,1)_k \times \M$ which are guaranteed to be ghost-free
by the method sketched above. 
A bosonic theory with target space 
${\rm SU}(1,1)_k \times \R^d$ is ghost-free
for any $d$ provided one chooses the rather bizarre-looking 
value $k = (26-d)/(23-d)$.
A more natural way to achieve $c=26$ in the bosonic case is to combine 
the compact and non-compact WZW models, taking 
${\rm SU}(1,1)_k \times {\rm SU}(2)_{\ell} \times \M^{20}$ 
with $k = \ell + 2$.
Finally, there is a very natural supersymmetric version of this,
${\rm SU}(1,1)_k \times {\rm SU}(2)_k \times \M^4$ which has 
$\hat c = 10$ from (\ref{ncc}). 
Notice that the compact and non-compact super WZW models 
appear here with the same
level. This is exactly the combination of WZW models which appears in the
simplest example of the AdS/CFT correspondence mentioned in the
introduction.\cite{M,MS,CT}

The arguments sketched above guarantee only 
that the string theory has no ghosts at the free level. 
To demonstrate consistency of the 
interacting theory it would be necessary to show that
crossing-symmetric amplitudes can be defined whose fusion rules close
among the ghost-free representations. 
For the case of SU(2) the fusion rules close amongst the unitary
representations, but it is not clear  
why this should be so for the ghost-free representations of the 
$\SU$ theory. This is an interesting topic for future work. 

\section{ Comments}

Finally, we return to the equivalence between 
type IIB string theory on ${\rm AdS}_3 \times S^3 \times \M^4$ and
a certain superconformal field theory in two dimensions.\cite{M} 
On the string side this involves a 
RR background of $Q_1$ D1-branes and $Q_5$ D5-branes, 
while the corresponding CFT has as its target space a symmetric
product of $k$ copies of $\M^4$ with $k \sim Q_1 Q_5$.
The correspondence should hold when both $Q_1$ and $Q_5$ are large. 
It was then pointed out\cspace\cite{MS} that 
since there are only finitely many chiral primary states in the CFT,
there must be a corresponding ``stringy exclusion principle''
which forbids certain states from appearing in the string spectrum. 

Under an S-duality transformation, this RR background 
becomes a conventional NS/NS background
of $Q_1$ fundamental strings and $Q_5$ NS5-branes.\cite{CT} For
$Q_1=1$ a perturbative string analysis of the type we have described 
here is valid, and the supersymmetric version of the 
restriction on representations (\ref{ncsl}) 
turns out to reproduce the same constraint as 
the stringy exclusion principle. 
Some caution is required in making this comparison, however, 
since in the first instance the AdS/CFT correspondence 
is expected to hold only if $Q_1$ is large.
Nevertheless, because of U-duality, the analysis for 
$Q_1\!=\!1$ will have implications for the case $Q_1$ large
and therefore the domains of validity of the 
constraints may in fact overlap.
It would be interesting to check this in more detail.

\vfill

\section*{Acknowledgments}
We thank Peter Goddard, Michael Green, Juan Maldacena, 
Hugh Osborn and Arkady Tseytlin for discussions and comments.

JME is grateful to the organizers of the Trieste Conference on Super 
Five-branes and Physics in $5{+}1$ Dimensions for the invitation to
speak and to the ICTP for warm hospitality. The research of JME is
supported by a PPARC Advanced Fellowship.
MRG is grateful to Jesus College and Fitzwilliam College, Cambridge
for financial support.

\end{document}